\documentclass{article}

\usepackage{PRIMEarxiv}

\usepackage[utf8]{inputenc}
\usepackage[T1]{fontenc}
\usepackage[numbers,square]{natbib}
\usepackage{url}
\usepackage{booktabs}
\usepackage{array}
\usepackage{tabularx}
\usepackage{longtable}
\usepackage{amsfonts}
\usepackage{nicefrac}
\usepackage{microtype}
\usepackage{lipsum}
\usepackage{fancyhdr}
\usepackage{graphicx}
\usepackage{pdfpages}
\usepackage{adjustbox}
\usepackage{xcolor}
\usepackage{xspace}
\usepackage{tikz}
\usepackage{float}
\usepackage{listings}
\usepackage[hidelinks]{hyperref}
\usepackage{cleveref}
\usetikzlibrary{arrows.meta,positioning,fit,calc}
\graphicspath{{media/}}
\usepackage{outlines}
\newcommand{\ecl}{\textsc{EconCSLib}\xspace}
\newcommand{\cslib}{\textsc{CSLib}\xspace}
\newcommand{\mathlib}{\textsc{Mathlib}\xspace}
\newcommand{\lean}{Lean\xspace}
\newcommand{\eclstatusdate}{July 2, 2026\xspace}
\newcommand{\eclpublicpapers}{24\xspace}
\newcommand{\eclformalizedpapers}{20\xspace}
\newcommand{\eclnonauthorformalizedpapers}{6\xspace}
\newcommand{\eclpublicpartials}{4\xspace}
\newcommand{\eclpubliclinesofcode}{986,391\xspace}

\newcommand{\eclauthorformalizedpapers}{14\xspace}

\newcommand{\eclprivateprogresspapers}{3\xspace}

\makeatletter
\newcommand{\DeclareTextualCiteAuthor}[2]{\@namedef{ecl@citet@author@#1}{#2}}
\renewcommand{\citet}[2][]{
  \@ifundefined{ecl@citet@author@#2}{
    \if\relax\detokenize{#1}\relax\citep{#2}\else\citep[#1]{#2}\fi
  }{
    \@nameuse{ecl@citet@author@#2}~
    \if\relax\detokenize{#1}\relax\citep{#2}\else\citep[#1]{#2}\fi
  }
}
\makeatother
\DeclareTextualCiteAuthor{bei2026econcslib}{Bei et al.}
\lstdefinestyle{lean}{
  basicstyle=\ttfamily\footnotesize,
  columns=fullflexible,
  keepspaces=true,
  breaklines=true,
  showstringspaces=false,
  xleftmargin=1em,
  frame=single,
  rulecolor=\color{black!20},
  literate={ℝ}{{$\mathbb{R}$}}1 {≠}{{$\ne$}}1 {→}{{$\to$}}1 {∃}{{$\exists$}}1 {∀}{{$\forall$}}1 {≤}{{$\le$}}1 {∧}{{$\land$}}1
}
\newcolumntype{P}[1]{>{\raggedright\arraybackslash}p{#1}}
\newcolumntype{Y}{>{\raggedright\arraybackslash}X}

\title{\ecl: AI-Assisted Lean Formalization for Economics \& Computation Research}
\date{\today}

\author{
  Nikhil Garg \\
  Cornell Tech \\
  Operations Research \& Information Engineering \\
  \texttt{ngarg@cornell.edu} \\
}

\begin{document}
\maketitle

\begin{abstract}
This paper presents \ecl, a \lean 4 library and workflow for formalizing research papers in applied modeling fields such as Economics and Computation, with language-model assistance. The goal of \ecl is to enable researchers to formalize their papers in \lean without knowing \lean themselves. The central design principle is a human-AI-\lean formalization workflow: an LLM writes \lean code, \lean checks formal statements and proofs, and both humans and LLM-as-judge processes can verify that the paper's statements were translated into \lean correctly. We develop agent skills, human-facing reporting, a review dashboard, and auditing procedures to support this workflow.

The current public repository contains \eclformalizedpapers formalized papers and \eclpublicpartials partially formalized papers, along with shared libraries for probability (including stochastic processes), auctions, matching markets, social choice, and graph tools, totaling \eclpubliclinesofcode lines of \lean code. To our knowledge, we are also among the first applied math researchers to systematically pursue \lean formalization of one's own publications in the process of building such a community library.

We welcome users and contributors to the project. The library and workflow are available at \url{https://github.com/nikhgarg/EconCSLib}, with corresponding project webpage at \url{https://gargnikhil.com/EconCSLib/}.

\end{abstract}


\section{Introduction}
Lean is a formal verification language for mathematics. If a statement is written and compiled in Lean, then the \lean statement is correct up to its assumed premises (subject to \lean's correctness). Formal verification holds great promise in guaranteeing the correctness of research papers with mathematical components. However, writing Lean code is tedious and difficult, and the libraries available in Lean have historically concentrated in pure mathematics. Thus, many areas of applied mathematics have not yet begun to be formalized, and formalization is out-of-reach for most researchers.

\textit{Auto-formalization}, through language models, is changing this.\footnote{The cost is non-trivial but manageable. Formalizing one paper takes about a full week's quota on a GPT Pro with Max tier account, which as of this writing costs \$200/month.} As of early 2026, the frontier large language models---e.g., GPT 5.5 Pro---became capable of writing Lean code at a level good enough to perform non-trivial verification of research: given a paper, translating its definitions, theorem statements, and proofs into Lean. Such writing of Lean code is arguably the ideal application of capable but error-prone AI: as long as the theorem statements and definitions are correctly translated from the intended statement (and this may be reasonably quickly checked by a human who can read Lean), the Lean \textit{compiler} verifies that the proof code (potentially hundreds of thousands of lines) correctly proves those translated statements -- no human needs to check these lines.\footnote{Of course, one needs to trust the Lean kernel/typechecker and its surrounding toolchain. In contrast, while frontier LLMs can directly find \textit{errors} in mathematical arguments, they cannot by themselves \textit{verify} that a statement is correct, without a language such as \lean; an LLM not finding an error is not sufficient evidence that there are no errors.} This human-AI-\lean workflow leverages the particular strength of each agent.
However, LLM agents cannot conduct trustworthy verification out-of-the-box, and starting from scratch for each paper is prohibitively wasteful.

This paper presents \ecl{} -- an AI-generated Lean library and corresponding LLM skills and a human validation interface -- to formalize Economics \& Computation research. Our goal is to put auto-formalization within reach of  researchers in the field for their own papers, without needing to learn \lean. \ecl tackles two key challenges: (a) building shared library infrastructure, encoding commonly used concepts across papers; (b) developing an auto-formalization workflow and auditing tools, to increase autonomy and trust that a paper has actually been formalized.

\ecl is organized as follows. The central workflow is to verify a specific research paper using a language model agent: given a paper's \LaTeX{} source or PDF, the agent extracts the mathematical statements (definitions, lemmas, theorems) and formalizes them in \lean, following the paper's given proofs to the extent possible. Papers from the community share many components, and so the workflow also builds a shared library: an individual paper's formalization can make use of this library, and the formalization agent ``elevates'' a statement from a paper's code to the shared library when it determines that it is likely to be useful for other papers.

This \lean code is supplemented with automated workflows and LLM skills to (1) aid AI agents in formalizing papers, by maintaining shared lessons across papers; and (2) automatically audit via code that each paper's formalization follows best practices. These skills and workflow are being actively updated as we formalize papers, with the goal to get close to full-automation of paper formalization. Most importantly, each paper's formalization contains (3) a set of artifacts to help \textit{humans} understand the formalization status and verify that the paper's statements have been translated correctly into \lean. These artifacts include a Directed Acyclic Graph (DAG) visualizing the paper's mathematical statements and current formalization status (see \Cref{fig:dag-excerpts} for the DAG associated with a fully formalized paper); a post-formalization report that documents the process such as whether the \lean proofs needed to deviate from the paper's approach or add additional assumptions, or whether any mistakes were found (see \Cref{app:validation-reports}); and a dashboard in which a human can verify (and log) whether the paper's statements were correctly translated into \lean (see \Cref{fig:verification-dashboard}). This last component, the dashboard to verify translation, is further supported by an LLM-as-judge module. These artifacts are automatically generated and maintained by the formalization agent.

At present, the library has been used to fully formalize \eclformalizedpapers papers; these include \eclauthorformalizedpapers papers by the author \citep{garg2019your,garg2021designing,liu2021testoptional,garg2021driver,peng2024reconciling,greenwood2024useritem,peng2025nofree,dong2024discretization,ma2025balancing,garg2019designing,dong2026capacity,garg2026combating,deshpande2024optimal,deshpande2026simpler}, and \eclnonauthorformalizedpapers other prominent papers from the EconCS community \citep{gale1962college,roth1982economics,goldberg2001competitive,mehta2007adwords,edelman2007internet,kleinberg2021algorithmic}. Another \eclpublicpartials papers are \textit{partially} formalized in the public repository, with the remaining formalization depending on infrastructure beyond the current scope of \ecl and not yet available in other libraries \citep{lehmann2002truth,lipton2004approximately,liu2024quantifying,garg2019iterative}; two require computational-complexity results, one requires Poisson process and stopping-time infrastructure, and one requires stochastic subgradient descent convergence formalization. The shared library's modules currently include those for probability and optimization, matching markets, social choice theory, and auctions and mechanisms, among others. We have especially focused on developing continuous probability tooling -- as commonly used in EconCS and Operations Research -- but which is especially challenging in \lean.

I stress that this project is experimental and early days: while formal auto-verification of mathematical research papers -- aided by language models -- is clearly valuable, it is unclear what role shared libraries like \ecl, and \ecl in particular, will play. Furthermore, while we have built tools to help humans verify the paper statements' translations into \lean code, this step is ultimately limited by having contributors with the requisite \lean expertise to do so. Finally, and relatedly, it is unclear how to organize \ecl as an open source software project to which anyone can contribute. One potential ultimate goal would be to allow others to use the workflow to formalize their own papers of interest, and then contribute its formalization and potential library-level components back to \ecl. Another goal could be to then, in turn, use this library as a basis for novel AI-generated proofs for EconCS papers.

\paragraph{Get started.} To get started in formalizing your own paper, clone the repository and open an LLM agent tool (I use Codex with GPT 5.5 in xhigh thinking mode). Give the agent the paper link, and ask it to formalize the paper using the skill and workflow in the repository. And please feel free to reach out as you do so. The repository is available at \url{https://github.com/nikhgarg/EconCSLib}, with project webpage at \url{https://gargnikhil.com/EconCSLib/}.

\paragraph{Contributions and organization.} Overall, this paper makes three contributions. First, it introduces a public Lean 4 library for Economics \& Computation, with reusable library infrastructure. Second, it presents a paper-oriented formalization workflow in which each research paper has a compact paper-facing Lean interface, a dependency DAG, a validation report, and dashboard- and LLM-as-judge-based translation review. Our formalization agent skills integrate lessons from our own formalization and best practices from the literature, with the goal of near-fully-automated formalization. Third, it reports an initial formalization case study, illustrating both the promise and the bottlenecks of author-led AI-assisted formalization. To our knowledge, we are also among the first applied math researchers to systematically pursue \lean formalization of one's own publications in the process of building such a community library.

The rest of the paper is organized as follows. \Cref{sec:related-work} discusses related work, and especially the substantial recent work on LLM-based autoformalization in \lean. \Cref{sec:architecture} describes the system architecture: shared library, paper-specific artifacts, scripts, skills, and review dashboard. \Cref{sec:current-status} summarizes the current formalization status and library coverage, alongside next immediate steps. \Cref{sec:discussion} discusses various aspects of auto-formalization and lessons learned.

\section{{Related Work}}
\label{sec:related-work}

\paragraph{\lean ecosystem and formalization projects}
\ecl builds on the \lean ecosystem \citep{deMoura2015Lean,deMouraUllrich2021Lean4}, and the mature, human-written, community-maintained mathematical library \mathlib \citep{mathlib2020}. \mathlib and similar projects emphasize that coordination and community workflows are as important as individual development in such projects \citep{vanDoorn2020MaintainingMathlib,baanen2025GrowingMathlib,bolan2025EquationalTheories}.

Domain-specific \lean projects now exist across several applied mathematics fields -- these include in high energy physics, chemical physics, and scientific computing~\citep{toobySmith2024HepLean,bobbin2024ChemicalPhysicsLean,scilean}.  \cslib is the closest analogue for computer science \citep{barrett2026CSLib}, aiming to formalize core computer science theory concepts; we expect \ecl to be a downstream user of \cslib{} -- for example, we are waiting to finish formalizing several papers until computational complexity libraries are more mature. We note that these projects are largely (but not exclusively) human-written, and aim to formalize central \textit{concepts} as opposed to research papers in a specific field.

We aim to build a workflow for human-AI-Lean formalization, and apply this workflow to formalize EconCS papers while building a library. Because of language model progress, substantial recent work has started on this path in other domains \citep{murphy2024LeanEuclid,wang2026M2F,gloeckle2026AutomaticTextbookFormalization,rammal2026formalizing,douglas2026FormalizationQFT,ilin2026VlasovMaxwellLandau,urban2026130k}, including workflows to formalize scientific papers \citep{meadows2026formalscience,ren2026merlean}. Zhang, Lee, and Liu \citep{zhang2026statistical} formalize core statistical learning concepts in Lean, using a Human-AI workflow. Many of these projects share workflow components with ours, including LLM generation of Lean code, compiler-guided proof repair, and human-facing tools for checking whether informal statements have been faithfully translated into formal statements. We note that many (but not all \citep{ren2026merlean,douglas2026FormalizationQFT,ilin2026VlasovMaxwellLandau}) of these existing projects focus on formalizing the textbook canon of a field, as opposed to a research-paper driven approach.

To our knowledge, \ecl is among the first public Lean projects to make systematic author-led formalization of research papers the organizing principle for a \lean
library in an applied area. The \mathlib Initiative \citep{formalFrontier2026} aims to bring autoformalization closer to the math research frontier, and is developing a specification to make such autoformalization more scalable and useful. We will adopt such a specification and its best practices.

Finally, we note that formalization has a longer history, including in game theory, cf. \citep{bagnall2017library}. Lean has also grown beyond formalizing mathematical results, to formalizing neural network-based software at runtime \citep{george2026TorchLean}, cryptographic code in Rust \citep{klaus2026rust}, and \lean itself \citep{carneiro2024lean4lean}.

\paragraph{Concurrent work: \citep{bei2026econcslib}} Independently and concurrently to this project, \citet{bei2026econcslib} also develop a \lean library entitled \ecl; both projects share substantial commonalities in aiming to ease \lean formalization of EconCS ideas and research, through the development of a library with concepts that are shared across papers. Both projects also make use of LLM tooling to support this goal.

The projects are highly complementary rather than redundant, stemming from a difference in philosophy.\footnote{I thank the authors of \citet{bei2026econcslib} for discussion contributing to this note. We have also discussed potentially more direct collaboration or integration in the future.} \citet{bei2026econcslib} have a tighter human-in-the-loop workflow aimed at correct and clean shared common abstractions -- a human \lean expert encodes the major definitions and statements, with proof support by an LLM. This enables, for example, precisely stating open problems in the language of the domain library. Our project, on the other hand, has a more automated workflow for formalizing research papers, with human-in-the-loop tooling aimed at ensuring that research paper statements are translated correctly; shared library concepts are an outcome of paper-focused development, and we do not pursue human curation or validation of the shared library concepts. In other words, the two projects are on opposing ends of a speed--human curation tradeoff: \citet{bei2026econcslib} requires humans who are experts in \lean and EconCS, yielding a human-curated and validated shared library; our project, in contrast, aims to make formalization of research papers accessible to researchers who do not know \lean, but in turn produces substantial supporting library code that has not been human-validated.

In the future, we hope to be able to use the curated concepts from the library of \citet{bei2026econcslib} in our project, and researcher use of our project may help guide which concepts are most important for humans to curate in the project of \citet{bei2026econcslib}. Our \ecl project already preferentially uses results from \mathlib and \cslib (which are human curated) when available, for analogous reasons.

\paragraph{Autoformalization and theorem proving methods} AI-assisted theorem proving and autoformalization is a growing methodological field. Autoformalization asks (language) models to translate informal mathematics into formal language \citep{wu2022autoformalization,weng2025autoformalization,mensfelt2026towards}; multiple benchmarks now measure progress in the field \citep{zheng2021minif2f,azerbayev2023ProofNet,patel2026mathatlas}. Research considers how to best build autoformalization pipelines, including incorporating RAG, RL, and fine-tuning \citep{xie2025fmc,lu2025automated,wu2026stepfun,huang2025formarl,shebzukhov2026improving}. Others build richer workflows, such as how to create proof blueprints, repair proofs using compiler feedback, improving translation of natural language statements into \lean statements, creating dependency graphs, and Human-AI interfaces \citep{zhu2026leanarchitect,ospanov2026apollo,jana2025proofbridge,wang2025aria,cabral2025proofflow,yanahama2026lean,lu2025formalalign,gao2025herald,ying2024lean}. \ecl incorporates some of these lessons into its skills and workflows (in particular, an LLM agent went through many of the above papers and integrated lessons into repository skill file).

Language models can go beyond formalizing human-written proofs into Lean -- they can now generate novel proofs~\citep{han2021proof,first2023baldur,lample2022Hypertree,jiang2022draft,ju2026automated,lin2025goedel,lin2025goedel2,polu2022formal,polu2020generative}; \citep{abouzaid2026first} seek to evaluate this ability while preventing leakage. Substantial work builds on top of Lean, such as to train or fine-tune language models to generate novel theorems \citep{yang2023leandojo,aniva2025pantograph,song2024towards,ren2025deepseek,xin2024deepseek,wang2025kimina,ma2026oprover,li2024hunyuanprover}. Our hope is that \ecl can be used by such provers to prove novel EconCS results. Lean is also used to provide supervision and generate synthetic data during training for LLM theorem provers \citep{chen2025seed,lin2025lean}.

\section{{\ecl Architecture}}
\label{sec:architecture}

\begin{table}[tbp]
\centering
\begingroup
\scriptsize
\setlength{\tabcolsep}{3pt}
\renewcommand{\arraystretch}{0.9}
\newcommand{\paperusersep}{;\allowbreak\space}
\caption{Reusable \ecl library components. The lines of code (LOC) counts are tracked \lean lines in the listed component files and submodules; the rows partition the public \texttt{EconCSLib/} Lean files, including folded umbrella imports and small related layers. Paper users are inferred from direct paper-local imports of the component or one of its submodules, so the column is a conservative dependency summary rather than a full transitive call graph over public artifacts.}\label{tab:library-components}
\begin{adjustbox}{max width=\textwidth,max totalheight=0.84\textheight}
\begin{tabularx}{\textwidth}{P{0.19\textwidth} Y P{0.08\textwidth} P{0.22\textwidth}}
\toprule
Library component & Content details & Lines of code & Paper users \\
\midrule
Foundations: finite math and graph tools & Core finite mathematics used across the formalizations: finite-choice distances, ranking and order conversions, finite-sum algebra, threshold and interval certificates, asymptotic estimates, and graph-cycle extraction for local-improvement arguments. & 15,941 & \citep{dong2024discretization}\paperusersep\citep{lipton2004approximately}\paperusersep\citep{garg2021driver}\paperusersep\citep{liu2021testoptional}\paperusersep\citep{garg2019iterative}\paperusersep\citep{kleinberg2021algorithmic}\paperusersep\citep{deshpande2024optimal}\paperusersep\citep{dong2026capacity}\paperusersep\citep{edelman2007internet}\paperusersep\citep{goldberg2001competitive}\paperusersep\citep{garg2019designing}\paperusersep\citep{peng2024reconciling} \\
Foundations: probability and stochastic processes & Probability infrastructure for recommendation, ratings, pricing, and ranking papers: finite PMFs and expectations, conditional kernels, Gaussian and random-utility comparisons, large-deviation and method-of-types certificates, stochastic dominance, order statistics, Markov chains, MDPs, CTMCs, Poisson processes, and renewal-reward models. & 76,470 & \citep{dong2024discretization}\paperusersep\citep{ma2025balancing}\paperusersep\citep{lipton2004approximately}\paperusersep\citep{garg2021driver}\paperusersep\citep{liu2021testoptional}\paperusersep\citep{liu2024quantifying}\paperusersep\citep{garg2019your}\paperusersep\citep{garg2021designing}\paperusersep\citep{garg2019designing}\paperusersep\citep{garg2019iterative}\paperusersep\citep{kleinberg2021algorithmic}\paperusersep\citep{edelman2007internet}\paperusersep\citep{goldberg2001competitive}\paperusersep\citep{peng2024reconciling} \\
Foundations: optimization, certificates, and complexity & Certificate-oriented optimization and complexity for paper-level endpoints: argmax, endpoint, and bisection principles; finite-search and output-check certificates; approximation and LP witnesses; binary-policy games; move-graph descent; strategic and choice equilibria; NP/ZPP consequence interfaces; and Yao-style lower-bound certificates. & 5,371 & \citep{dong2024discretization}\paperusersep\citep{mehta2007adwords}\paperusersep\citep{garg2021driver}\paperusersep\citep{liu2021testoptional}\paperusersep\citep{lipton2004approximately}\paperusersep\citep{garg2019iterative}\paperusersep\citep{kleinberg2021algorithmic}\paperusersep\citep{lehmann2002truth}\paperusersep\citep{deshpande2024optimal}\paperusersep\citep{deshpande2026simpler}\paperusersep\citep{garg2019designing}\paperusersep\citep{peng2024reconciling} \\
Social choice, rankings, voting, and fair division & Executable social-choice infrastructure: ranked ballots, next-active support counts, RCV/STV traces, quotas, coalition-safety lemmas, final-order structures, candidate-deletion reductions, Thiele-style proportionality, Mallows/Kendall ranking models, payoff interfaces, envy graphs, bounded-envy algorithms, and indivisible-goods fairness statements. & 23,773 & \citep{lipton2004approximately}\paperusersep\citep{garg2019your}\paperusersep\citep{garg2019iterative}\paperusersep\citep{kleinberg2021algorithmic}\paperusersep\citep{deshpande2024optimal}\paperusersep\citep{deshpande2026simpler}\paperusersep\citep{garg2026combating} \\
Auctions and mechanisms & Mechanism-design infrastructure for digital-goods, combinatorial, and position auctions: allocation and payment semantics, utility formulas, DSIC truthfulness, VCG-style welfare maximization, benchmark-competitive auctions, single-minded set packing, greedy mechanisms, and critical-value certificates. & 12,881 & \citep{goldberg2001competitive}\paperusersep\citep{lehmann2002truth}\paperusersep\citep{edelman2007internet} \\
Online algorithms and regret & Online allocation infrastructure centered on AdWords and platform learning: matching/allocation state machines, primal-dual accounting, competitive-ratio certificates, regret interfaces, and hooks for randomized lower-bound arguments. & 4,374 & \citep{mehta2007adwords}\paperusersep\citep{ma2025balancing} \\
Matching markets and admissions & Stable matching and admissions infrastructure: one-to-one and many-to-one assignments, blocking-pair stability, deferred-acceptance invariants, applicant/proposer optimality, quota admissions, strategic application cutoffs and payoffs, and two-group policy objective comparisons. & 3,924 & \citep{gale1962college}\paperusersep\citep{roth1982economics}\paperusersep\citep{dong2026capacity}\paperusersep\citep{liu2021testoptional} \\
Ratings and recommender systems & Ratings and recommendation infrastructure: Bayesian binary and ordinal signal models, posterior-mean rating formulas, prior-weighted updates, monotonicity and correction lemmas, Thompson-sampling entrypoints, exposure/allocation policies, classwise fairness constraints, top-k recommendation surfaces, policy averaging, and accuracy/diversity trade-off statements. & 3,186 & \citep{ma2025balancing}\paperusersep\citep{garg2021designing}\paperusersep\citep{garg2019designing}\paperusersep\citep{greenwood2024useritem}\paperusersep\citep{peng2024reconciling} \\
\bottomrule
\end{tabularx}
\end{adjustbox}
\endgroup
\end{table}

\begin{figure}[p]
\centering
\includegraphics[width=\textwidth,height=0.82\textheight,keepaspectratio]{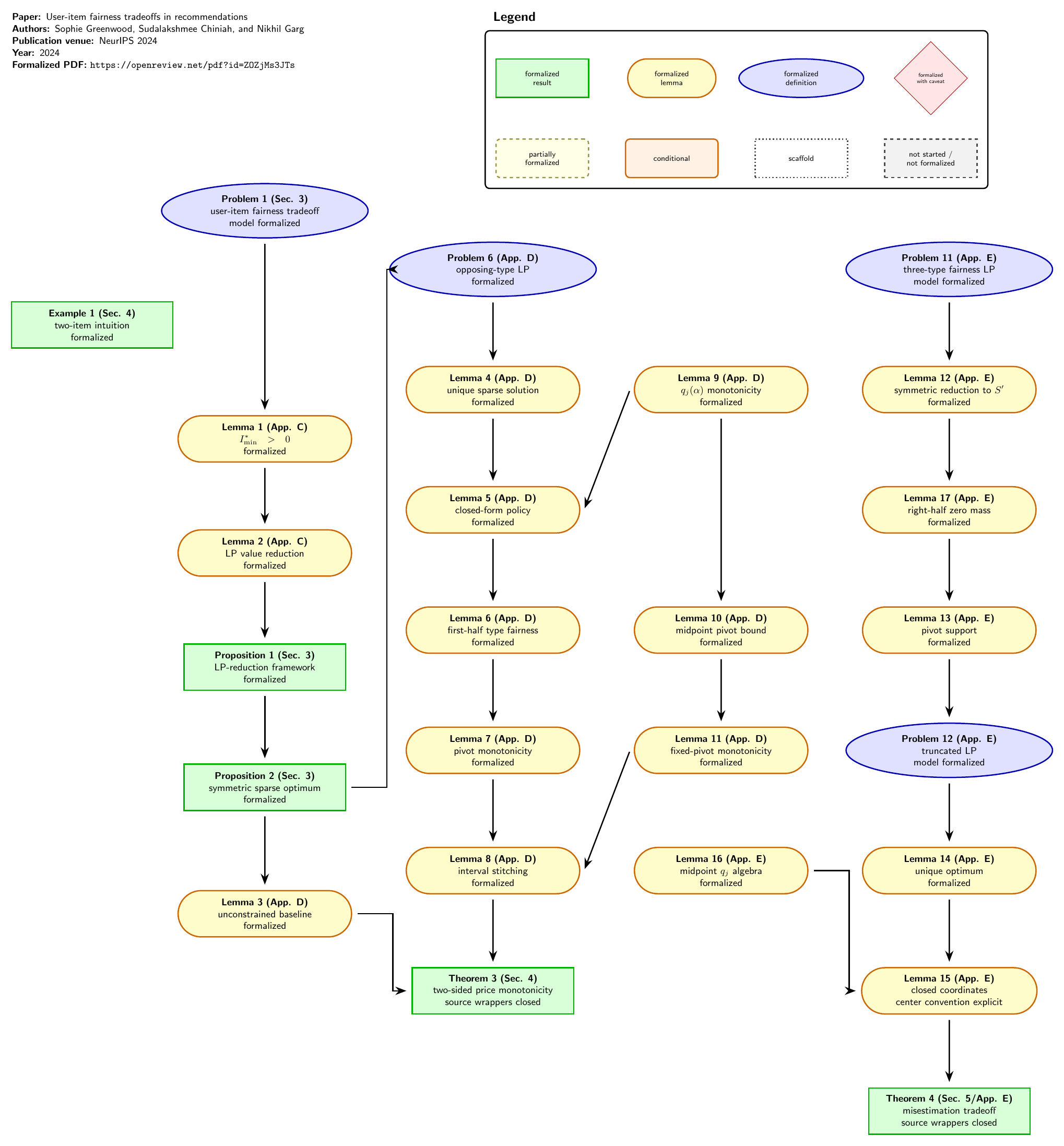}
\caption[Full paper statement dependency DAG]{Full paper statement dependency DAG for \citep{greenwood2024useritem}. The DAG is auto-generated at the beginning of a paper's formalization process, and it is kept up-to-date so that a human can quickly understand formalization status. This DAG shows that the paper has been completely formalized.}
\label{fig:dag-excerpts}
\end{figure}

\begin{figure}[tbp]
\centering
\includegraphics[width=\textwidth]{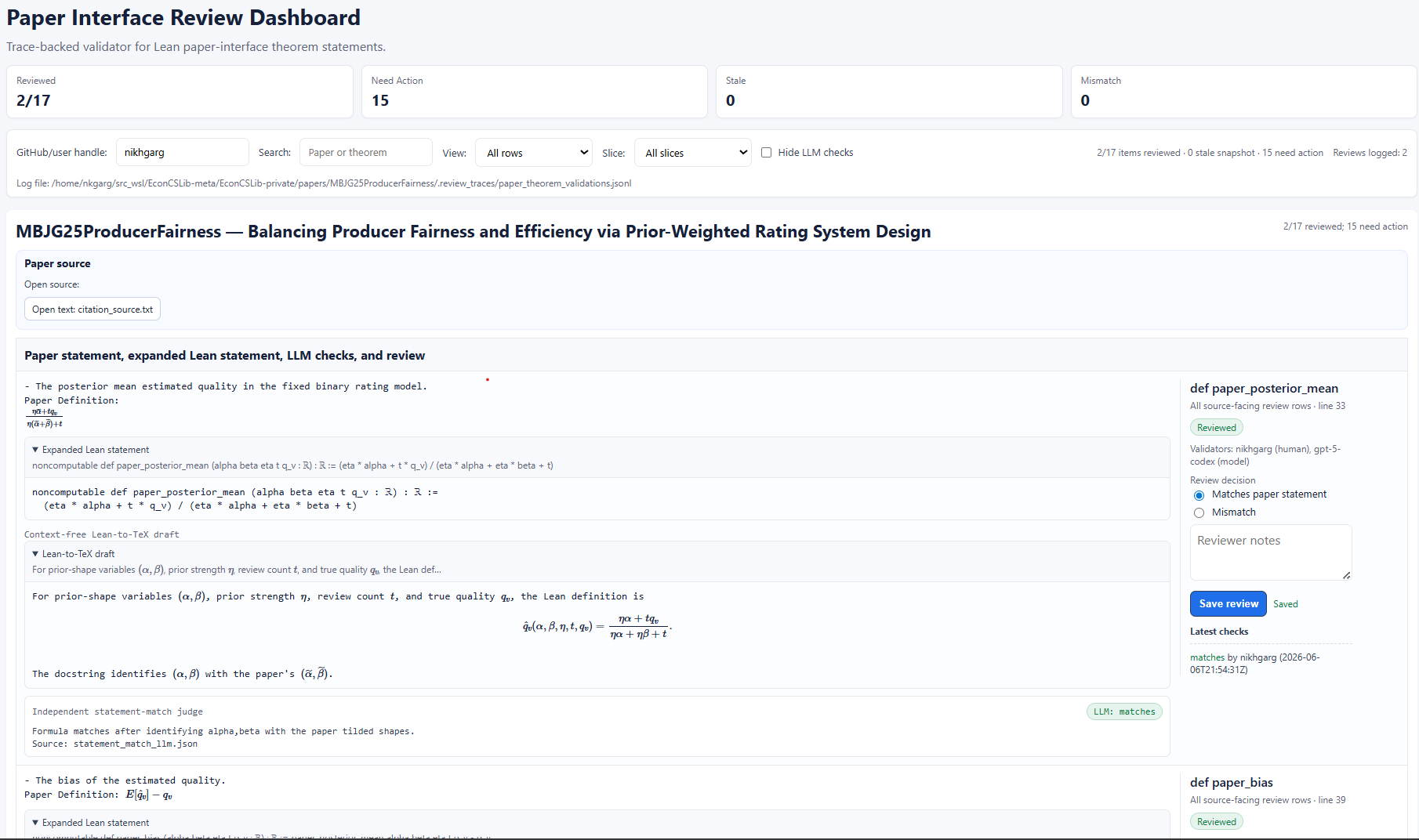}
\caption[Paper-interface review dashboard]{Screenshot of the paper-interface review dashboard for \citep{ma2025balancing}. For each paper mathematical statement, the dashboard shows the paper's \LaTeX{} statement and \lean statement, so that a human can verify that these match. The dashboard further shows an AI-generated Lean-to-TeX draft (generated without any context for the original paper)---ideally, this should match the original paper, and may be helpful for the human to verify translation accuracy. Finally, there are human review controls for recording whether the formal statement matches the source claim, which get logged into the repository.}
\label{fig:verification-dashboard}
\end{figure}

Here, we describe the system components in more detail.

\subsection{\lean code} \ecl has two layers of \lean code. The reusable \textit{library} layer, in \texttt{EconCSLib/}, contains paper-independent, common definitions and results for areas such as probability, optimization, matching, auctions, online algorithms, recommender systems, rankings, and fair division. The current library components are listed in \Cref{tab:library-components}. The \textit{research} layer, in \texttt{papers/}, has one folder per research paper; each folder is meant to audit and formalize a specific research paper preserving source notation, theorem numbering, formal statements, and proof approach. Each paper folder also contains the human-facing validation artifacts as described below.

\subsection{Statement translation validation and human-in-the-loop workflow} The \lean code is not sufficient on its own. Researchers have to be able to verify that the \lean code actually has formalized what the paper intended it to, be able to monitor progress, and understand the process's outcome.

\subsubsection{Human and agentic validation of translation} While the \lean compiler verifies that a given \lean statement can be derived from the axioms, it cannot confirm that the paper's intended mathematical statement was correctly translated to the \lean statement. This task must be done outside \lean, and is where formalization errors are most common. Notably, a human does not need to validate every \lean statement was correctly translated, just the paper facing statements -- the \lean compiler can do the rest. Thus, while the \lean proof files can extend to over a hundred thousand of lines, the workflow enforces that each paper has a compact \texttt{PaperInterface.lean} file that just contains these paper-given definitions and theorem statements (but not proofs). The translation verification task is thus to validate that the statements in these files correctly translate the intended statements from the paper.

\ecl pursues two complementary approaches to validating this translation: (1) LLM-as-judge validation; (2) a dashboard to support human understanding and validation. Both approaches save persistent logging documenting which statements' translations have been validated. 

\paragraph{LLM validation} The first validation layer is itself automated, via an LLM-as-judge approach. This validation has three components: \textit{coverage}, \textit{translation accuracy}, and \textit{holistic audit}. The first two components are at the individual statement level: each mathematical statement from the paper is extracted into a table, and mapped to specific \lean statements; then, an LLM-as-judge approach is used to validate that each source statement is covered by some \lean statement, and that the \lean statement is a correct translation of the mathematical statement.\footnote{First, an LLM (without any context on the source paper) is asked to translate \textit{from} the \lean code \textit{back} to \LaTeX. Then, a second LLM (also without any context) is asked to compare the \LaTeX translation to the source statement in the paper, and judge whether they are equivalent.} The third component, as a holistic audit, focuses on the connections between statements as opposed to a statement-by-statement approach---in early formalization campaigns, a failure mode was that the statement-by-statement approach made errors due to semantic drift between statements. This workflow can (and does) raise when the judge LLM is ``uncertain'' about a match, flagging it for human review. 

This pipeline can be run anytime, and is currently designed to run both at the beginning of a paper formalization and at the end. At the beginning, it can be used to iterate on the intended \lean targets until translation-verified \lean statements are found. Python tooling, supported by GitHub Continuous Integration (CI) checks, enforces that this pipeline has been run and is passing before a paper can be declared formalized.

We note that, at the moment, all validations are done by a codex (sub-)agent itself, and so self-preferencing or correlated errors \cite{kim2025correlated} concerns exist (that the judge agent makes the same mistakes that the original translation and \lean-writing agents do). 

\paragraph{Human validation dashboard} As illustrated in \Cref{fig:verification-dashboard}, in the dashboard, each Lean statement is presented alongside the original paper statement (visualized from the original \LaTeX{} if available) and the LLM translation and verdict as described above. The human can then mark whether the \lean statement matches the intended paper statement, which is stored in the repository. Thus, the human work in formalizing the paper is analogous to a human checking that the theorem statement is correctly written down, as opposed to checking the proof directly. As we discuss below, this step is the key bottleneck for autoformalization, and where \ecl is most lagging. 

~\\ Outputs from each of the above validations are stored persistently in the repository itself, including the validator identity, timestamp, and any comments -- provenance of all judgements (github username, or LLM model with timestamp) as well as hashes of the validated \lean statements are included in the documentation. We note that errors and imprecision in this step are the most common issues found in the formalization process so far: since it is not programmatic but rather human and using a language model, subtle issues can emerge such as a statement being assumed in a \lean function statement as opposed to proven in its proof. Over time, we have focused on improving the LLM-as-judge and automated checks (in Python) for this task; we foresee this audit tooling being improved over time, as failures are detected and patched. 

\subsubsection{Other human-facing artifacts and documentation.} Two other human-facing artifacts enable the human-in-the-loop process and steering of the formalization agent.

(1) \textbf{DAG with concise result formalization status.} As illustrated in \Cref{fig:dag-excerpts}, the LLM agent generates and then maintains a DAG of the paper's named mathematical statements (e.g., definitions, lemmas, theorems); while the paper is not fully formalized, this DAG contains the status of each result. Monitoring this DAG is a key way that the human can track progress and steer the agent toward certain results; status updates documented in the DAG also give insight on where the LLM may be stuck, and need guidance.

(2) \textbf{Final human-facing validation report.} The workflow produces a \texttt{FINAL\_VALIDATION\_REPORT.md} that complements the DAG in describing the formalization process, in a human digestible form. Its components include: the formalization status of each result, whether the agent was able to follow the paper's proof approach or needed to deviate (either due to an error, or a missing library-level result that it did not build), and what if any additional assumptions were needed to prove the result. See \Cref{app:validation-reports} for examples; for example, the validation report for \citep{ma2025balancing} reports that a theorem statement's strict inequality required an extra assumption implicit in the paper (that Bernoulli success probabilities $p$ were bounded away from 0 and 1). This report also tracks the current status of the LLM-as-judge and human translation verification.

\subsection{Agent ``skill'' and automation scripts to autoformalize papers} Finally, as we're formalizing papers, we ask the agent to log lessons for future agents. LLM agents use markdown files to document such learnings; these markdown files are called Skill files, and are often shared online. The goal here is to create the self-orchestration scaffolding so that an LLM, with no other context, can receive a paper link and follow a workflow to formalize it. These learnings are split into two components:

(1) ``Workflow'' skills and auditing scripts.  These skills document and automate workflows for starting and then ending a paper: for example, they document that to start a paper, the agent should acquire the paper pdf and \LaTeX{} source from arXiv if available; then, it should extract mathematical statements and create a formalization plan and the DAG. After finishing a paper, the agent should check that the \texttt{PaperInterface.lean}, DAG, and \texttt{FINAL\_VALIDATION\_REPORT.md} files follow best practices. Other  skill components include advice to: keep the paper interface reviewable; record proof-route deviations immediately; and upstream reusable lemmas only when a second paper is likely to need them.
Non-markdown scripts, such as those in Python, also check for things like ``sorry,''  a \lean command that can take the place of a proof to allow a file to compile without a real proof; templates for each of the human-facing artifact files are also included.

(2) ``Proof'' skill files: In addition to the above ``Workflow'' skills, we are creating ``Proof best practices'' skills -- both for overall formalization suggestions and guardrails, and for specific types of mathematical claims. Within each library component, files document component file structure and suggestions on various proof approaches. For example, for probability-heavy papers, the skill suggests that a default interpretation for many statements should be an "up to measure 0" result, as opposed to statements that are true pointwise everywhere. These skills also document common proof patterns needed to turn standard statements (such as that a certain type of limit converges) into formal Lean statements.

~\\Developing this skill in the process of formalizing papers -- and especially the audit tooling as discussed above -- has been a primary output of the project, and where human feedback has been most encoded. This skill has been developed through three main avenues. First and foremost, we instruct Codex to encode feedback and lessons from our interactions into the skill file, especially when it comes to course-corrections in the formalization or documentation process. Second, this feedback-to-skill process was then itself automated, using the approach of \citet{guo2026skilldisco}: a Codex agent was instructed to analyze all its past sessions and then extract durable lessons from its traces and our interactions into the skill file. Third, an LLM agent extracted workflow lessons from the literature on autoformalization, and hardened skill patterns where appropriate \citep{wang2026M2F,cabral2025proofflow,zhu2026leanarchitect,wang2025aria,first2023baldur,ospanov2026apollo,ren2026merlean,meadows2026formalscience,lu2025formalalign}.

\section{{Current Formalization Progress}}
\label{sec:current-status}

As of \eclstatusdate, the public repository contains \eclpublicpapers paper formalizations: \eclformalizedpapers fully formalized papers and \eclpublicpartials partial formalizations, totaling \eclpubliclinesofcode lines of \lean code. \Cref{tab:paper-status} details these papers and their status. These include \eclauthorformalizedpapers formalized papers co-authored by the author of this paper, and \eclnonauthorformalizedpapers other prominent EconCS papers, including those selected from the test-of-time awardees of the ACM EC conference. This approach -- formalizing both the author's papers and classic papers -- balances (a) developing the formalization workflow using results with which the author is familiar and (b) building out the shared library with results most likely to be useful in formalizing other EconCS papers.

In addition, \eclprivateprogresspapers non-public paper formalizations are active in our private repository.

I stress that these papers are formalized in the following sense:
\begin{itemize}
\item The \lean code compiles, and so Lean has checked the correctness of the translated statements, given the stated assumptions and definitions. 

\item Our LLM-as-judge and post-formalization audit workflow concludes that the \lean code fully \textit{covers} the mathematical statements in the paper, that they have been translated correctly, and that no additional assumptions not in the paper (except where documented) are made. 

\item While the human verification dashboard exists, it has only been sparsely used to verify the translation fully captures the paper statement, and that no extra assumptions have been made. 

\end{itemize}
 In other words, for these papers, we have completed the ``AI agent'' and ``\lean compiler'' parts of the Human-AI-\lean process. However, we have not yet completed the expensive ``human'' cost of this process, where a human with some knowledge of reading \lean verifies the paper-facing statements have been correctly translated. A big open question for this project, and similar such projects, is how they will complete this translation validation process with human checkers. In the meantime, we are relying on an LLM-as-judge approach to aid in this process. We foresee that this will be the default workflow for most intended users of \ecl.

\section{{Discussion}}
\label{sec:discussion}
\begin{table}[tbp]
\centering
\begingroup
\small
\setlength{\tabcolsep}{4pt}
\renewcommand{\arraystretch}{1.0}
\caption{Codex token usage for \ecl formalization sessions across two measured machines through June 5, 2026, when the public repository contained 13 paper formalizations, with about 5 more in progress. Codex was asked to go through its own session logs and count tokens, deduplicating counts across resumed sessions and subagents.}\label{tab:codex-token-usage}
\begin{adjustbox}{max width=\textwidth}
\begin{tabularx}{\textwidth}{Y >{\raggedleft\arraybackslash}p{0.22\textwidth} >{\raggedleft\arraybackslash}p{0.15\textwidth} >{\raggedleft\arraybackslash}p{0.17\textwidth}}
\toprule
Quantity & Logged tokens & Credits & Approximate cost\textsuperscript{\dag} \\
\midrule
Uncached input tokens & 665,103,654 & 83,138 & \$3,326 \\
Cached input tokens & 22,054,285,696 & 275,679 & \$11,027 \\
Output tokens, including reasoning & 68,771,979 & 51,579 & \$2,063 \\
Reasoning output tokens & 24,456,589 & -- & Included in output \\
\midrule
\textbf{Total} & \textbf{22,788,161,329} & \textbf{410,396} & \textbf{\$16,416}  \\
\bottomrule
\end{tabularx}
\end{adjustbox}
\vspace{0.25em}
\parbox{0.98\textwidth}{\footnotesize
\textsuperscript{\dag}\textit{Cost calculation.} Approximate cost uses the OpenAI Codex GPT-5.5 token-based rate card as viewed June 1, 2026: 125 credits per million uncached input tokens, 12.5 credits per million cached input tokens, and 750 credits per million output tokens. Fast mode would multiply the credit cost by about a factor of 2, and was often used. We note that we did not spend this in raw dollars -- these tokens were included as part of a GPT Pro subscription, which costs \$200/month (and were provided to the author for free).
}
\endgroup
\end{table}

While the project is still in its infancy, there are several lessons that can be drawn and open questions important for its long-term contributions.

\paragraph{Human monitoring and steering} The workflow is not yet fully automated -- not quite at the vision in which we can hand an agent the paper pdf and receive full \lean code verification. However, it is getting closer, and the Codex agent can now work for many hours (e.g., overnight or all day) on a paper without human intervention or feedback.

While I have not written a single line of \lean code, for the early papers I actively monitored the Codex windows that are writing the proofs and intervene when it seems to get stuck. For example, it might start generating large proof sections designed to circumvent needing to use an assumption that is implicit in the paper. (In one case during verification of \citep{garg2021driver}, it constructed large proof sections because it did not realize that the paper's assumptions imply that the denominator of a given equation cannot be 0). The agent sometimes also stops at not proving the full paper statements from raw primitives but instead assumes an intermediate step. Several times, the agent also stalled at trying to prove pointwise statements where the paper (implicitly or explicitly) proved an ``up to measure 0'' statement. \Cref{app:codex-feedback-workflows} details some of this back-and-forth, based on an analysis of session logs.

Over time, as the agent formalization skill has been developed, these common traps have been documented, automated audit tooling has been built, and autonomy has increased substantially, especially given the audit procedures, the LLM-as-judge workflow, and in the development of human-facing artifacts. 

\paragraph{Human translation verification -- is it feasible?} The primary open question is how far human verification should go for a paper. In theory, the goal of a library like \ecl is that downstream paper consumers can ``trust'' a library as having provided correctly translated statements. For example, a \lean paper-facing statement for \cite{ma2025balancing} is as follows, referring to \texttt{EconCSLib.Statistics.JensenConvex}, a \ecl library function defining convexity.
\begin{lstlisting}[style=lean]
theorem paper_facing_theorem3_2_squared_bias_convex_in_quality
    {alpha beta eta t : ℝ}
    (hden : eta * alpha + eta * beta + t ≠ 0) :
  EconCSLib.Statistics.JensenConvex
      (fun q => paper_squared_bias alpha beta eta t q)
\end{lstlisting}

A human in the dashboard can verify that this statement matches the paper's statement, but someone needs to verify that what the paper means by convexity is correctly implemented in \texttt{EconCSLib.Statistics.JensenConvex}. Ideally, not \textit{every} human verifier of a paper that uses convexity needs to do this validation.

More generally, most intended users of \ecl will not know enough \lean code to be able to verify that it has correctly translated their paper statements, nor will they wish to invest the effort to verify the translations. For example, the function that defines Thompson sampling looks as follows.
\begin{lstlisting}[style=lean]
def paper_facing_thompson_sampling_mechanism
    {V : Type*} [Fintype V] [DecidableEq V] [Nonempty V]
    (belief : PMF (V → ℝ))
    (policy : PMF V) : Prop :=
  ∃ tie_breaker : (V → ℝ) → V,
    (∀ (profile : V → ℝ) (v : V),
      profile v ≤ profile (tie_breaker profile)) ∧
    policy = belief.bind (fun profile => PMF.pure (tie_breaker profile))
\end{lstlisting}

Whether this precisely matches what we mean by Thompson sampling is not apparently obvious to someone who is not familiar with \lean code (including the author of this paper).

For this reason, we have implemented the LLM-as-judge workflow to validate paper translations. This may be insufficient for high-stakes validations; whether it is sufficient for an intended use case such as validating proofs of submitted papers to conferences is an open, sociological question.

\paragraph{Human vs AI verification in Lean more generally} \cite{bessis2026theoremEconomy} discusses early tensions between the \mathlib community -- which is pursuing painstaking, clean, human verification of pure mathematics papers -- and various math AI startups, that are leveraging the work of this community without, potentially, giving enough back. There are substantial questions for \textit{how} a library like \ecl can give back to \mathlib and other human-curated repositories: ultimately, while verified for correctness via the Lean compiler, \ecl is still a form of ``AI Slop'' -- its code is bloated and messier than the analogous human project would be.

Besides aesthetics, there are also questions of \textit{intermediate} translation correctness: while paper-facing statement translations can be confirmed by a human (though this itself has costs, as discussed above), not every intermediate library statement can be. Thus, it may not be clear what exactly a library actually contains until it is needed by an agent to prove a particular statement in another paper (for example, the convexity definition issue discussed above). Such ambiguity likely prevents any integration of AI-generated libraries with human-curated \lean projects.

As discussed above, human verification of translated statements is the primary contrast with the \ecl project of \citet{bei2026econcslib}. Their project only includes statements in the shared library that have been curated and validated by an expert in \lean (and EconCS as needed). This results in a more trusted shared library (they may not face this ``convexity'' issue), but cannot be built in as automated a manner. A ``best of both worlds'' approach would be to preferentially use formalized concepts from trusted libraries when available; our \ecl project agent skill specifies such preferencing.

\paragraph{Token cost} The process is \textit{expensive} in terms of tokens. This project is only possible on an academic budget because of the heavy subsidization by OpenAI for tokens within Codex. Formalizing one paper approximately takes a full week's quota on an OpenAI Pro subscription on the Max tier, the highest tier available. \Cref{tab:codex-token-usage} quantifies the overall approximate token usage on this project so far (calculated by summing the token usage across measured sessions in the project folders). For the first 13 fully formalized papers (and about 5 in progress), this added up to about \$16,000 (or more if fast mode usage is counted) if paid for directly on a per-token API cost basis as opposed to via a Pro subscription on the Max tier. Reducing this cost, such as through better agent scaffolding, is an important direction for future work; we also hope that as the shared library is further developed (especially for continuous probability), this cost will decrease.

\paragraph{Dependence on other libraries and standard results} Just like the EconCS community depends on mathematical advances in other fields, \ecl depends on mathematical libraries in \lean, notably \mathlib; this project would not be possible without the years of careful human work to develop such libraries. Notably, some required results are not yet available in standard libraries, and so we are left with the following choices: (1) pause on formalizing a paper fully until those results are available, such as by the \cslib library; (2) declare ``victory'' by \textit{assuming} those (potentially standard) results; (3) develop the necessary results ourselves in our \lean library; (4) prove our final desired statement another way which does not need the result. For example, for \citep{lehmann2002truth} and \citep{lipton2004approximately}, we have gone with choice (2), assuming standard computational complexity results. For \citep{greenwood2024useritem}, the agent avoided invoking a result on the number of basic feasible solutions in a linear program by proving the necessary statements in a more paper-local, ad hoc fashion. For several other papers, the agent proved (and elevated to our probability library) statements about Gaussian random variables and Continuous Time Markov Chains that may one day exist in an upstream \lean library.

One notable characteristic is that \lean is better suited for discrete math arguments, as opposed to the continuous, measure-theoretic probability arguments needed for many EconCS or Operations papers. However, LLM-generated formalization now makes formalizing papers with substantial probability feasible, if not easy. For example, the formalizations of \cite{liu2021testoptional} and \cite{garg2021driver} each took well over 100,000 lines of \lean code, much of it to define and formalize measure theoretic equilibria equations (for example, that a given strategy was optimal up to measure 0). Our hope is that the probability and equilibria-related components being developed for \ecl make formalizing such papers more efficient in the future.

\paragraph{Errors and hand-waviness in published papers} This project raises several ``sociological'' questions about correctness in applied modeling EconCS papers. While we have not yet found any ``fatal'' mistakes in a paper, we have found several minor ``bugs'' that can be patched, unspoken assumptions, or true theorem statements whose proofs were nevertheless incorrect, requiring new proofs (that the LLM agent was able to find). Furthermore, some proof steps, even if correct, are under-specified to the point that it takes the LLM agent substantial work to try to complete the step. For example, the agent was able to formalize our paper \citep{dong2024discretization}; however, a proof step was seemingly underspecified, and the final validation report documents the following:
\begin{quote}
Theorem 1 proof deviation. The paper's continuous source-transformation proof sketch is underspecified at the measurable transformation step and in the multiclass $S_b,S_d$ mass accounting. Lean proves the same paper-facing bound directly from calibration, and formalizes the source-transformation route using an explicit real-coordinate sweep with coordinate pushforward and pullback. This is a proof-strategy deviation, not a theorem-statement change.
\end{quote}

In the past, such issues have largely gone under the radar, and the community generally believes that most such errors if they exist are inconsequential if they can be repaired with minor edits. We are now at a time where such errors can be found at scale using LLMs, and not all papers will be successfully formalized by \lean, which by definition does not permit such hand-waving. This raises several questions about process: if I as an author find such substantial hand-waving (or even an error that can be patched by an additional assumption or clarification), do I submit an erratum for journal publications? Post a note on arXiv? By some standards, such issues will be found in many papers, and then the community will be drowning in errata. What if I find such issues in someone else's paper? Is it reasonable to email them, or just post a verification report in a repository? While such issues regarding errors have always existed, we are now at a time that they can no longer slip under the radar.

\paragraph{Long term goals and roadmap} This project is in its extremely early days, and it is unclear how it can provide value to the community. However, it seems to be a project worth doing, both to verify the EconCS canon and published research papers. One can imagine a world, perhaps in the very near future (STOC`27 or EC`27 perhaps), in which every submitted paper comes with an AI-generated \lean verification of its statements. Libraries such as \cslib and \ecl aim to make such verification more efficient and in reach for every author. Similarly, many researchers may seek to verify all their own published papers, a process that I have begun for myself.

Of course, AI generation of novel proofs is also a rapidly advancing area; it may be the case that \lean libraries help in this process for EconCS papers. \citet{koren2026theorist} is building a toolkit for LLMs to help in the development and validation of applied models in economics. As more papers are verified, it may be that the library's progress makes this process easier, for both verification of human-written and AI-generated proofs.

One vision, that hopefully will come about, is that people can focus on the creativity, conceptual insights, and real-world usefulness components of applied modeling -- while AI helps most with ensuring correctness and helping prove generality.

I welcome contributions by others for the library and to aid in this effort.

\section*{Acknowledgments}
I thank Guanting Chen, Jiajun Ma, Xiaohui Bei, and members of the Garg lab for helpful discussions. The full formalization work so far was completed through a single OpenAI Codex Pro subscription on the Max tier, provided for free by Arjun Seshadri at OpenAI. I am supported by NSF CAREER IIS-2339427, the William T Grant Foundation Scholars Award, NASA, and Google, Mastercard, and Amazon research awards.

\begingroup
\scriptsize
\setlength{\tabcolsep}{1.8pt}
\renewcommand{\arraystretch}{1.08}
\setlength{\LTleft}{0pt}
\setlength{\LTright}{0pt}
\begin{longtable}{@{}P{0.33\textwidth} P{0.095\textwidth} P{0.13\textwidth} P{0.065\textwidth} P{0.34\textwidth}@{}}
\caption{Current formalization status of papers in \ecl.}\label{tab:paper-status}\\
\toprule
Paper info & Lean proof & Human translation & Lines of Code & Note \\
\midrule
\endfirsthead
\caption[]{Current formalization status of papers in \ecl{} (continued).}\\
\toprule
Paper info & Lean proof & Human translation & Lines of Code & Note \\
\midrule
\endhead
\midrule
\multicolumn{5}{r}{\scriptsize Continued on next page}\\
\endfoot
\bottomrule
\endlastfoot
\emph{College Admissions and the Stability of Marriage}. D. Gale and L. S. Shapley; American Mathematical Monthly, 1962~\citep{gale1962college}. & Formalized & 0/7 reviewed & 388 & This only uses a few lines of code as its infrastructure has largely been elevated to the shared matching library. \\
\emph{The Economics of Matching: Stability and Incentives}. Alvin E. Roth; Mathematics of Operations Research, 1982~\citep{roth1982economics}. & Formalized & 0/29 reviewed & 8,931 &  \\
\emph{Competitive Auctions and Digital Goods}. Andrew V. Goldberg, Jason D. Hartline, and Andrew Wright; SODA, 2001~\citep{goldberg2001competitive}. & Formalized & 0/30 reviewed & 14,624 & Formalizes the SODA paper; Theorem 8.2 uses the refined monotone-auction wording from the journal version~\citep{GOLDBERG2006242}. \\
\emph{AdWords and Generalized Online Matching}. Aranyak Mehta, Amin Saberi, Umesh Vazirani, and Vijay V. Vazirani; Journal of the ACM, 2007~\citep{mehta2007adwords}. & Formalized & 0/43 reviewed & 13,711 &  \\
\emph{Internet Advertising and the Generalized Second-Price Auction}. Benjamin Edelman, Michael Ostrovsky, and Michael Schwarz; American Economic Review, 2007~\citep{edelman2007internet}. & Formalized & 0/26 reviewed & 133,338 &  \\
\emph{Designing Optimal Binary Rating Systems}. Nikhil Garg, Ramesh Johari; AISTATS / PMLR 89, 2019~\citep{garg2019designing}. & Formalized & 0/56 reviewed & 85,780 &  \\
\emph{Who is in Your Top Three? Optimizing Learning in Elections with Many Candidates}. Nikhil Garg, Lodewijk Gelauff, Sukolsak Sakshuwong, Ashish Goel; HCOMP, 2019~\citep{garg2019your}. & Formalized & 0/17 reviewed & 35,055 &  \\
\emph{Designing Informative Rating Systems: Evidence from an Online Labor Market}. Nikhil Garg, Ramesh Johari; Manufacturing \& Service Operations Management 23(3), 2020~\citep{garg2021designing}. & Formalized & 0/15 reviewed & 7,027 &  \\
\emph{Algorithmic Monoculture and Social Welfare}. Jon Kleinberg and Manish Raghavan; PNAS, 2021~\citep{kleinberg2021algorithmic}. & Formalized & 0/49 reviewed & 65,646 &  \\
\emph{Test-optional Policies: Overcoming Strategic Behavior and Informational Gaps}. Zhi Liu and Nikhil Garg; EAAMO, 2021~\citep{liu2021testoptional}. & Formalized & 0/23 reviewed & 125,744 &  \\
\emph{Driver Surge Pricing}. Nikhil Garg and Hamid Nazerzadeh; Management Science, 2022~\citep{garg2021driver}. & Formalized & 0/36 reviewed & 142,948 &  \\
\emph{Optimal Strategies in Ranked-Choice Voting}. Sanyukta Deshpande, Nikhil Garg, Sheldon H. Jacobson; arXiv:2407.13661, 2024; working paper~\citep{deshpande2024optimal}. & Formalized & 0/65 reviewed & 54,410 &  \\
\emph{Reconciling the Accuracy-Diversity Trade-off in Recommendations}. Kenny Peng, Manish Raghavan, Emma Pierson, Jon Kleinberg, and Nikhil Garg; The ACM Web Conference, 2024~\citep{peng2024reconciling}. & Formalized & 0/42 reviewed & 52,018 & Proposition 2's printed finite bound appears to miss a factor of 2; Lean proves the corrected finite bound, which is sufficient for the asymptotic 1/2-homogeneity result. \\
\emph{User-item fairness tradeoffs in recommendations}. Sophie Greenwood, Sudalakshmee Chiniah, and Nikhil Garg; NeurIPS, 2024~\citep{greenwood2024useritem}. & Formalized & 0/48 reviewed & 46,174 &  \\
\emph{A No Free Lunch Theorem for Human-AI Collaboration}. Kenny Peng, Nikhil Garg, Jon Kleinberg; AAAI, 2025~\citep{peng2025nofree}. & Formalized & 0/15 reviewed & 2,030 &  \\
\emph{Addressing Discretization-Induced Bias in Demographic Prediction}. Evan Dong, Aaron Schein, Yixin Wang, and Nikhil Garg; PNAS Nexus, 2025~\citep{dong2024discretization}. & Formalized & 0/41 reviewed & 26,133 &  \\
\emph{Balancing Producer Fairness and Efficiency via Prior-Weighted Rating System Design}. Thomas Ma, Michael S. Bernstein, Ramesh Johari, and Nikhil Garg; ICWSM, 2025~\citep{ma2025balancing}. & Formalized & 10/27 reviewed; 2 uncertain & 680 & Strict variance decrease is formalized with the explicit interior-quality assumption 0 < q\_v < 1. \\
\emph{Capacity Constraints Make Admissions Processes Less Predictable}. Evan Dong, Nikhil Garg, Sarah Dean; AAAI-26 published version, DOI 10.1609/aaai.v40i45.41179~\citep{dong2026capacity}. & Formalized & 0/66 reviewed & 11,536 &  \\
\emph{Combatting Gerrymandering with Ranked Choice Voting: an Experimental Analysis of Multi-member Districts in the United States}. Nikhil Garg, Wes Gurnee, David Rothschild, David Shmoys; Operations Research, 2026, DOI 10.1287/opre.2024.1167~\citep{garg2026combating}. & Formalized & 0/19 reviewed & 9,398 &  \\
\emph{Simpler Than You Think: The Practical Dynamics of Ranked Choice Voting}. Sanyukta Deshpande, Nikhil Garg, Sheldon H. Jacobson; Journal of Computational Social Science, 2026~\citep{deshpande2026simpler}. & Formalized & 0/50 reviewed & 20,736 &  \\
\emph{Truth Revelation in Approximately Efficient Combinatorial Auctions}. Daniel Lehmann, Liadan Ita O'Callaghan, and Yoav Shoham; Journal of the ACM, 2002~\citep{lehmann2002truth}. & Partially formalized & 0/39 reviewed & 7,582 & Greedy approximation, truthfulness, and Theorem 6.1 reductions are formalized. Full formalization requires computational complexity results that are out of scope. \\
\emph{On Approximately Fair Allocations of Indivisible Goods}. Richard J. Lipton, Evangelos Markakis, Elchanan Mossel, and Amin Saberi; ACM EC, 2004~\citep{lipton2004approximately}. & Partially formalized & 0/48 reviewed & 80,496 & Sections 2 and 4 are fully formalized. Section 3 has query/descent/rounded-search support. The PTAS/FPTAS runtime layer needs reusable fixed-dimension IP complexity infrastructure. \\
\emph{Iterative Local Voting for Collective Decision-making in Continuous Spaces}. Nikhil Garg, Vijay Kamble, Ashish Goel, David Marn, Kamesh Munagala; JAIR 64, 2019~\citep{garg2019iterative}. & Partially formalized & 0/47 reviewed & 23,466 & Partial formalization: the remaining intentional boundary isolated to the SSGM convergence theorem. Theorem 3 is proved as a constrained alternative in general and as the original statement under the explicit full-space condition. \\
\emph{Quantifying Spatial Under-reporting Disparities in Resident Crowdsourcing}. Zhi Liu, Uma Bhandaram, Nikhil Garg; Nature Computational Science, 2024~\citep{liu2024quantifying}. & Partially formalized & 0/27 reviewed & 18,540 & Full formalization requires a homogeneous Poisson process and stopping-time derivation. \\
\end{longtable}
\endgroup

\newcommand{\etalchar}[1]{$^{#1}$}

\clearpage
\appendix

\section{{Human Feedback Workflows}} \label{app:codex-feedback-workflows}

\textbf{Note:} The initial text for this appendix section was written by Codex, based on its session logs: about 30 distinct sessions from April 23 through June 6, 2026, with over 2,000 user prompts. The author performed an editing pass on the appendix, for substance, style, and concision, but much of the raw text originally generated by Codex remains. The main text was written by the author, with some supporting information provided by Codex.

This appendix summarizes the human steering workflow used to formalize individual papers and to develop the library itself.

\subsection*{Summary: Lessons for Human-AI Formalization}

The most productive feedback was both operational and mathematical.
The user did not only say whether a theorem was right; the user repeatedly
specified what artifact would make the claim reviewable. Examples include:
\texttt{PaperInterface.lean} as the statement checkpoint, the DAG as the
source-dependency checkpoint, \texttt{status.json} as the status source of
truth, and the final validation report as a one-page human audit rather than an
agent progress log.

The second lesson is that expert feedback should be compiled into the workflow.
Several recurring corrections became durable instructions: download and cache
the source PDF once; avoid broad documentation churn during proof loops; commit
at named theorem or stopping boundaries; keep generated tables tied to status
metadata; use official publication metadata; distinguish formalized, partial,
conditional, and caveated results; preserve human-written summaries; and keep
the main formalization skill mostly paper-agnostic, with paper-specific proof
recipes living in library notes or paper-local plans. This reduced repeated
mistakes across later papers.

The third lesson is that the human's role is especially important at the
translation boundary from informal paper mathematics to formal statements. The
agent can write and repair \lean{} proofs, but the user often supplied the
decisive question: does this theorem state the paper's claim, are the
certificates internally discharged, does a green DAG node depend on an unproved
assumption, and is the remaining boundary a missing proof or a missing reusable
library? Those questions are the main reason the project can expose partial
formalizations honestly while still making completed paper claims useful for
human readers.

\subsection*{A Feedback Loop for Paper Formalization}

The basic workflow was not a one-shot ``paper link to proof'' process. It was a
loop in which Codex proposed or implemented a formalization step, \lean{} checked the proof objects, and the user corrected the translation boundary, review surface, or repository process. Over time, those corrections were written into the repository skill file, status scripts, validation reports, and contribution workflow. In this sense, the user feedback did two jobs: it improved individual paper formalization, and it converted mistakes into durable instructions
for future agents.

The loop stabilized around several stages alongside the \lean{} work. First, the agent was told to get context
from the repository and the formalization skill, then inventory the paper's
definitions, lemmas, propositions, theorems, and displayed theorem-like claims.
Second, the agent built a source dependency DAG and a proof plan before or
alongside \lean{} work. Third, the agent formalized the reusable mathematics and
paper-facing wrappers, prioritizing theorem seams over a line-by-line encoding
of the PDF. Fourth, the agent exposed a small human-review surface:
\texttt{PaperInterface.lean}, a dependency DAG, a paper README, and a final
validation report. Fifth, the user audited whether those surfaces faithfully
matched the paper and whether any assumptions or certificates remained. Sixth,
the resulting status metadata regenerated the README, website, paper tables,
and public/private release artifacts.

\subsection*{How the Interaction Pattern Changed Over Time}

The earliest sessions were mostly persistence-oriented proof sessions. The user
gave broad goals such as fully formalizing a family of papers and repeatedly
pushed the agent to continue rather than stop at a local obstruction. This often
appeared as a queue of short steering commands: ``keep going,'' ``continue,''
``you do not have a time limit,'' ``just keep going until the paper is
formalized,'' and ``create sub-agents as needed.'' Over time this moved the
interaction from short supervised edits to longer autonomous loops, where the
agent could spend several hours on a proof seam, run targeted builds, repair the
mathematical route, and continue after resumptions without asking for each next
step. The main feedback at this stage was about effort allocation: commit less
often, avoid updating broad reports during every proof loop, cache source PDFs
instead of re-searching, and spend more time repairing imprecise informal proofs
outside \lean{} before returning to code; this feedback is now integrated into the skill.

A second phase shifted from proof production to reviewability. After the agent
claimed progress on paper theorems, the user asked what a human should inspect
to know that the paper was actually formalized. That changed the workflow from
``many compiling declarations'' to a smaller review contract:
\texttt{PaperInterface.lean} in source order, a dependency DAG containing the
paper's named results, and a final validation report written for a human rather
than for the next agent. Next, we focused on semantics of status and assumptions. The user pressed
on whether results were actually proven according to the source paper, or whether green DAG nodes depended on
unproved inputs. This produced stricter status language: completed paper-facing
claims are \texttt{formalized}; results that are not fully deduced from the basic \lean axioms are \texttt{partially
formalized}; and caveats are reserved for source-version corrections or extra
assumptions that remain in the final theorem.

The next phase was artifact and publication governance. Once the repository
was moving toward a public release and this paper, the feedback became more
about source-of-truth discipline: generated tables should come from
\texttt{status.json}, public pages should mirror the same metadata, unfinished
papers should stay private by default, bibliography entries should come from
official sources, and human-authored status summaries should not be overwritten
once marked as human-written or human-approved. A later consolidation phase sharpened the workflow. For example, post-formalization review became a stricter workflow: DAGs should contain the paper's numbered results, final reports should be human-facing; further, the dashboard and LLM-as-judge translation verification approach was developed. Reproducibility and measurement became explicit requirements: source PDFs and \TeX{} should be cached locally, generated status files should be updated only at meaningful boundaries, and token/cost accounting should deduplicate resumed sessions and subagent rollout logs.

\subsection*{Examples of User Feedback and Resulting Workflow Rules}

\begin{center}
\small
\begin{tabular}{P{0.23\textwidth} P{0.34\textwidth} P{0.33\textwidth}}
\toprule
Feedback type & Example user feedback & Resulting workflow rule or artifact \\
\midrule
Autonomy and persistence &
Queued commands such as ``keep going,'' ``continue,'' and ``you do not have a
time limit,'' with permission to think outside \lean{} and patch imprecise
informal proofs. &
Agents now work toward clean theorem or compile boundaries, use targeted builds,
and write handoff notes only at real stopping points. The skill explicitly says
not to stop at identifying a gap if the proof can be repaired. \\
\addlinespace
Proof-loop pacing &
Feedback such as ``focus on proving things,'' ``do not keep updating
\texttt{status.json},'' ``commit less,'' and ``update documentation only at
major milestones.'' &
During active proof work, agents should prefer theorem progress, local builds,
and proof-plan repair. Status JSON, DAG PDFs, skill notes, and commits are
batched at named proof or audit boundaries rather than every proof loop. \\
\addlinespace
Source inventory and DAG discipline &
Create a DAG for every named definition, lemma, theorem, proposition, and
corollary; each numbered result should be one node; green nodes must state the
actual result and mean that the paper-facing theorem is fully formalized. &
Each paper folder maintains a dependency DAG. The DAG is treated as a source
map, not a changelog of helper lemmas, and status colors must agree with the
paper-local \texttt{status.json}. \\
\addlinespace
Human-facing \lean{} interface &
The decisive review object should be a \lean{} file whose definitions and
theorems appear in paper order; if \lean{} checks it and a human confirms it
matches the paper, the paper is done. &
\texttt{PaperInterface.lean} became the canonical statement surface.
Implementation endpoints move to \texttt{ProofInterface.lean} or
\texttt{PostPaperAudit.lean} when they would make the review file too large. \\
\addlinespace
Validation report scope &
The final validation report is not a handoff log; it should be a short human
report: what was proved, what assumptions were added, what mistakes were found,
and whether the proof follows the paper. &
Completed papers get concise \texttt{FINAL\_VALIDATION\_REPORT.md} files.
Long theorem ledgers and status chatter are kept out of the human report. \\
\addlinespace
Status vocabulary and assumption boundaries &
``Formalized'' and ``verified in Lean'' should not be separate statuses. If a
paper-facing theorem still requires an external certificate, mark the paper
partial. &
The status vocabulary was standardized. Fully closed papers have status
\texttt{formalized}; papers such as
\citep{lehmann2002truth,lipton2004approximately} remain partial because the
remaining complexity/runtime infrastructure is outside the current library. \\
\addlinespace
Library extraction &
When proof infrastructure is reusable, elevate it to the shared library; e.g.,
matching and auction code should explain why some papers have few paper-local
lines. &
Reusable probability, optimization, matching, auction, online-algorithm, and
fair-division components were separated from paper folders. Paper-local code is
kept as source-shaped wrappers over shared infrastructure. \\
\addlinespace
Publication and provenance hygiene &
Public/private repositories should protect unfinished papers; public tables
should be generated from status JSON; BibTeX should come from official sources;
human-written summaries should not be overwritten. &
The project now has private-workflow contribution docs, generated
\texttt{human\_status.json} tables, GitHub Pages scaffolding, bibliography
hygiene rules, and review-status metadata for human-authored summaries. \\
\addlinespace
Token and cost accounting &
The user challenged naive counts that double-counted resumed sessions and
embedded parent transcripts inside subagent rollouts. &
The workflow now treats root-session history as the human-feedback record and
counts usage from each rollout's own turn boundary. Guardian/auto-review logs
are separated from billed interactive work unless explicitly being audited. \\
\bottomrule
\end{tabular}
\end{center}

\subsection*{Feedback on DAGs and Human-Facing Audits}

A recurring theme was that proof artifacts needed to be legible to a human
reviewer, not merely useful to the agent. For dependency DAGs, the user pushed
for source completeness and semantic accuracy: every named result should appear;
node labels should describe the paper claim rather than the implementation
helper family; green should mean the paper-facing theorem is fully formalized;
and a theorem should not be marked green if its visible inputs are still open.
The user also gave presentation-level feedback: avoid overlapping nodes and
arrows, do not duplicate a numbered proposition across several boxes, do not use
internal labels such as \texttt{lem:...} as the visible node name, and state the
mathematical content of a green theorem rather than merely saying that it is
closed. Empirical sections can be acknowledged in reports, but they do not
belong in proof DAGs unless the paper states a theorem-like empirical claim that
is being formalized. Venue metadata also belongs on the DAG: the formalized PDF
may be an arXiv version, but the public node header should still say, for
example, HCOMP 2019, AISTATS, or MSOM when that is the official publication
venue. This turned the DAG from a progress sketch into a human-auditable
dependency map whose status colors had to agree with the validation report and
\texttt{status.json}.

The same pattern applied to final audits. The user repeatedly separated three
documents that the agent had initially blurred: README/handoff notes for future
work, \texttt{PaperInterface.lean} for statement-level human review, and
\texttt{FINAL\_VALIDATION\_REPORT.md} for a short final audit. The report was
not supposed to be a theorem ledger, dashboard dump, or agent status transcript;
it should answer whether the paper is formalized, what assumptions or
certificates remain, whether the proof follows the paper strategy or uses a
cleaner route, and whether any source mistakes were found. This also led to
filtering large dashboard surfaces---for example, hundreds of rows in one paper were too
many for a human report---and to treating human summaries in generated tables as
review prose that should be preserved once marked human-written or
human-approved.

This feedback also affected dashboards. Human-review row counts should track
paper-facing definitions and named results, not every helper theorem produced
during implementation. A paper with dozens of rows may be correct if the source
has many named statements, but high row counts require an audit pass: shrink
\texttt{PaperInterface.lean} to source-facing statements, move proof endpoints
to \texttt{ProofInterface.lean}, and keep generated dashboards from creating an
inflated impression of the human audit burden.

\subsection*{Feedback on Assumptions and Status Language}

Several sessions were devoted to the difference between a caveat, a partial
formalization, and a source-theorem assumption. The user repeatedly asked
whether an apparently extra hypothesis was actually stated in the paper, whether
it could be derived from the paper's model, and whether the final theorem
should expose it. In one paper, this changed descriptions of several lemmas:
the issue was not that the result was conditional on an external certificate,
but whether the formal statement reflected a meaningful difference from the
source paper. In the ranking/rating papers, the same question arose for strict
rankings, finite support, threshold behavior, and Laplace-style limits. The
resulting workflow rule is to search the source first, and derive assumptions that
the source structure already implies.

\subsection*{Representative Case Studies}

\paragraph{Ranking and rating papers.}
The ranking and rating papers \citep{garg2019your,garg2021designing,garg2019designing} show the interaction between source reading,
large-deviation boundaries, and library extraction. The user first asked for
three large-deviation papers to be formalized in a sequence, starting with the
multi-candidate election paper \citep{garg2019your} but instructing the agent to produce library-level large deviations tooling useful for all three papers. For
the two rating-system papers \citep{garg2021designing,garg2019designing}, the user
pushed on whether a generic Laplace principle was truly required or whether the
paper's finite-support and strict-ranking hypotheses already removed the
apparent threshold obstruction.
That changed the plan from ``stop at a broad LDP boundary'' to ``derive the
finite-support convention used by the paper, close what can be closed, and
extract any genuinely reusable rating-model or finite-sum optimization lemmas to
the shared library.''

\paragraph{Auctions and digital goods.}
The auction formalization \citep{goldberg2001competitive,GOLDBERG2006242} illustrates how statements might be refined or clarified in published papers. \ecl formalizes the SODA version \cite{goldberg2001competitive} of the paper. However, the journal version \cite{GOLDBERG2006242} is referenced to interpret the statement for Theorem 8.2 -- the journal
paper refines this part of the preliminary presentation by introducing monotone
auctions and proving the upper bound for monotone truthful randomized auctions.

\paragraph{Complexity boundaries.}
The combinatorial-auction and fair-division papers \citep{lehmann2002truth,lipton2004approximately} show how the user can decide on the boundary between what should be proved versus assumed.

For the combinatorial-auction paper \citep{lehmann2002truth}, the
decision was: greedy approximation, truthfulness, and Theorem 6.1
reductions are formalized, while full formalization required computational
complexity results that are out of scope. For the fair-division paper
\citep{lipton2004approximately}, the proof boundary similarly stops at fixed-dimension integer-program runtime infrastructure.

\paragraph{Producer fairness.}
The producer-fairness paper \citep{ma2025balancing} demonstrates a different kind of expert feedback:
the user asked whether a caveat was really a paper assumption, a derived fact,
or an extra condition introduced by the formalization. Ultimately formalizing the paper required a trivial assumption (that Bernoulli success probabilities were not exactly 0 or 1) for the strict endpoint; this was documented as an additional assumption, not as a result caveat or incorrect statement.

\clearpage
\section{{Supplemental Information}}
\label{app:supplemental-information}
\label{app:dependency-dags}

\begin{table}[H]
\centering
\small
\begin{tabularx}{\textwidth}{@{}lY@{}}
\toprule
Label & Meaning \\
\midrule
\texttt{formalized} & The paper-facing claim is represented by a Lean statement and closed by a Lean proof from the paper's source-shaped assumptions. \\
\texttt{formalized with caveat} & The Lean proof closes, but a reader should account for a material qualification, such as an added non-paper assumption, a corrected or clarified source statement, or a documented paper issue. \\
\texttt{partially formalized} & The paper has a declared formalized subset, but at least one paper-facing result or required source layer remains outside the current Lean proof. \\
\bottomrule
\end{tabularx}
\caption{Formalization status labels used in paper summaries, validation reports, and status tables.}
\label{tab:formalization-status-labels}
\end{table}

\begin{table}[H]
\centering
\small
\begin{tabularx}{\textwidth}{@{}lY@{}}
\toprule
Label & Meaning \\
\midrule
\texttt{match} & Lean statement faithfully captures the paper-facing claim. \\
\texttt{uncertain} & Reviewer cannot decide without auditing a shared predicate, hidden convention, or source ambiguity. \\
\texttt{mismatch} & Lean statement differs materially from the paper claim. \\
\texttt{stale} & Review was made against an older statement hash. \\
\texttt{missing} & Source statement lacks a corresponding Lean statement. \\
\bottomrule
\end{tabularx}
\caption{Statement-translation audit labels used in the dashboard and LLM-as-judge summaries.}
\label{tab:statement-translation-labels}
\end{table}

\begin{figure}[H]
\centering
\includegraphics[width=\textwidth,height=0.86\textheight,keepaspectratio]{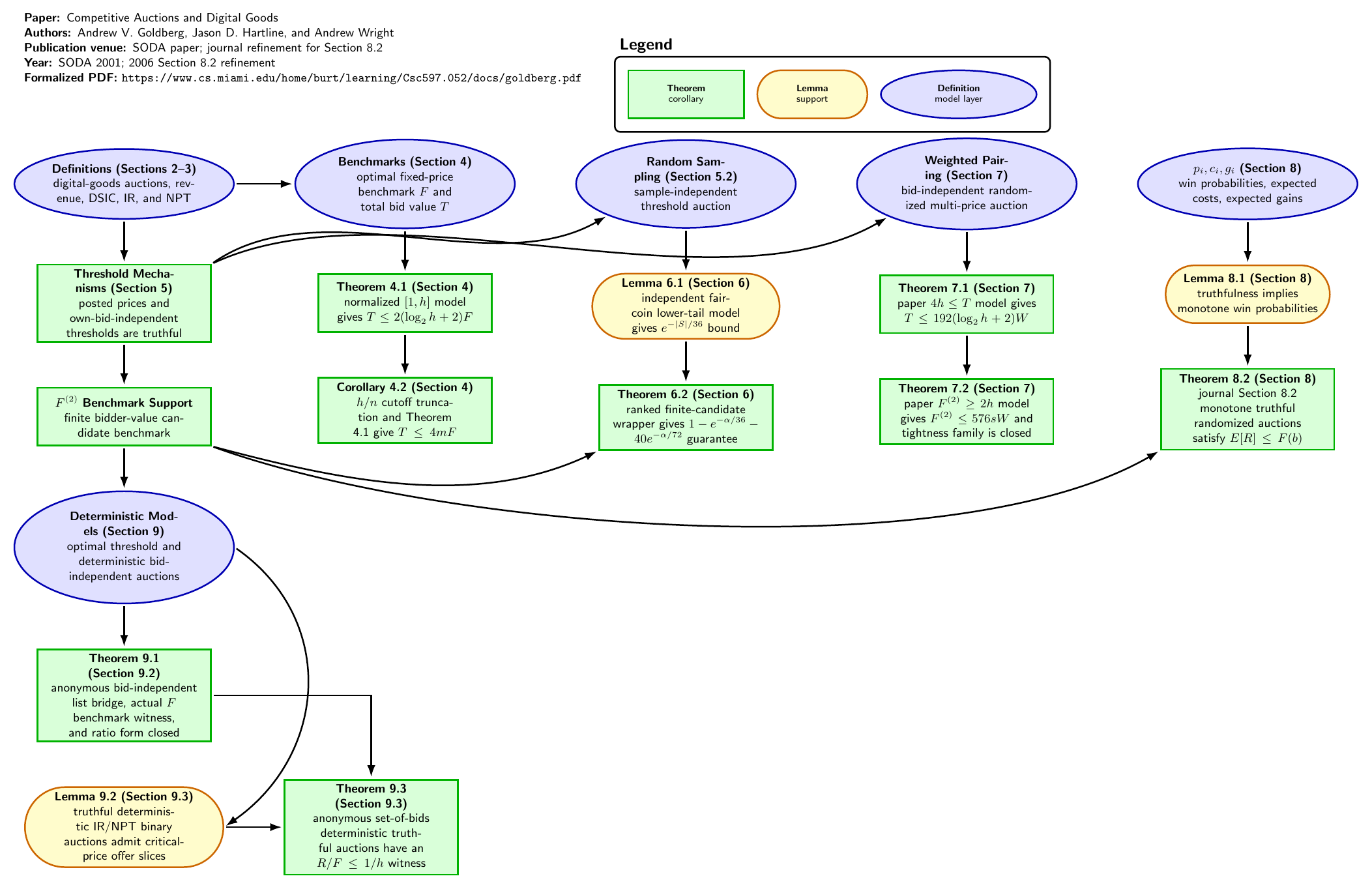}
\caption[Full dependency DAG for Competitive Auctions and Digital Goods]{Full dependency DAG for \citep{goldberg2001competitive}, \emph{Competitive Auctions and Digital Goods}. The DAG shows the SODA paper. Section 8.2 is checked against the later journal version's monotone-auction formulation of the revenue upper bound; the preliminary unrestricted wording is tracked only as source-version provenance.}
\label{fig:ghw01-dag}
\end{figure}

\clearpage
\includepdf[pages=1,pagecommand={\section{{Sample Validation Reports auto-generated by LLM workflow}}\label{app:validation-reports}}]{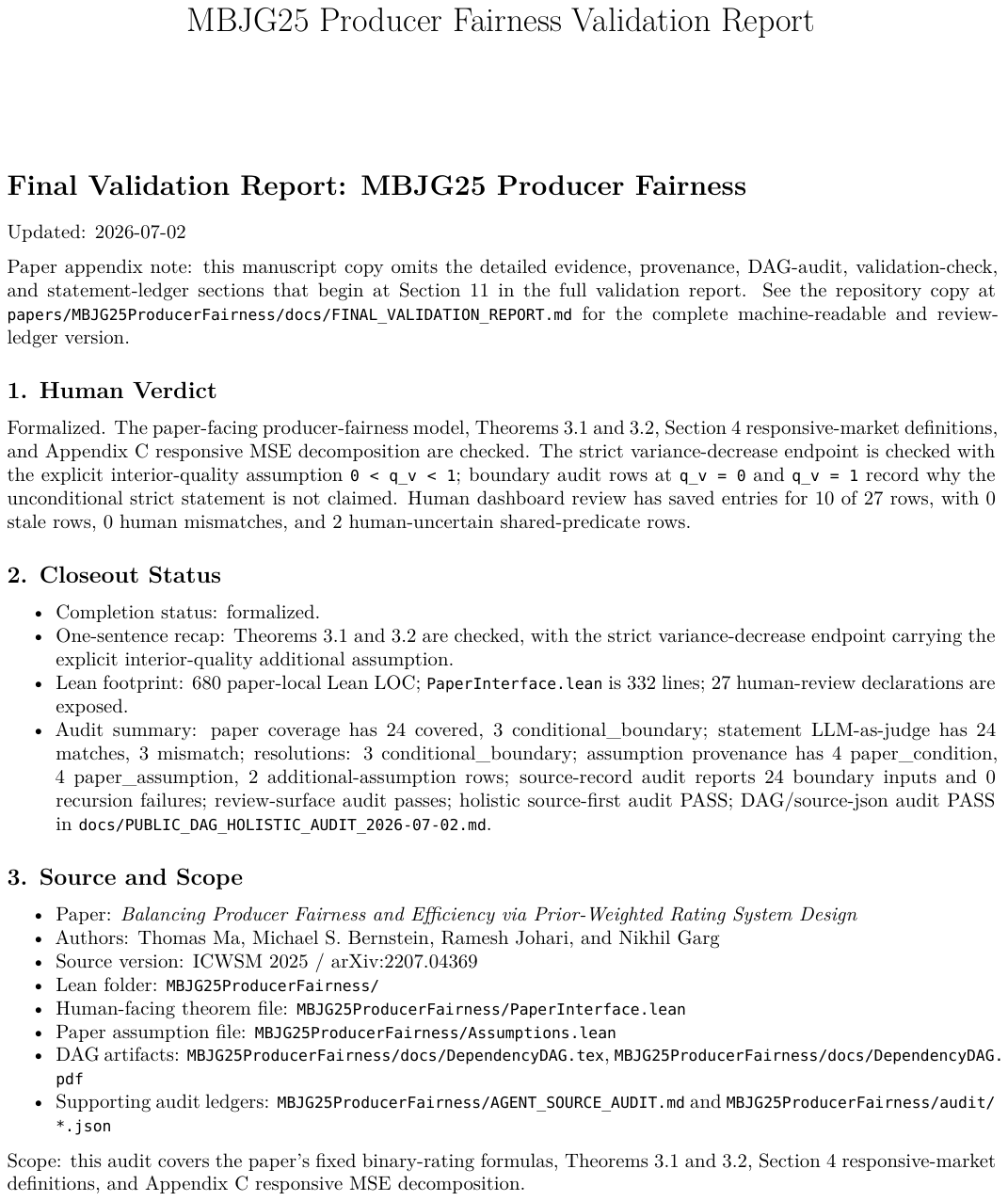}
\includepdf[pages=2-,pagecommand={}]{appendix/MBJG25_FINAL_VALIDATION_REPORT.pdf}

\includepdf[pages=-,pagecommand={}]{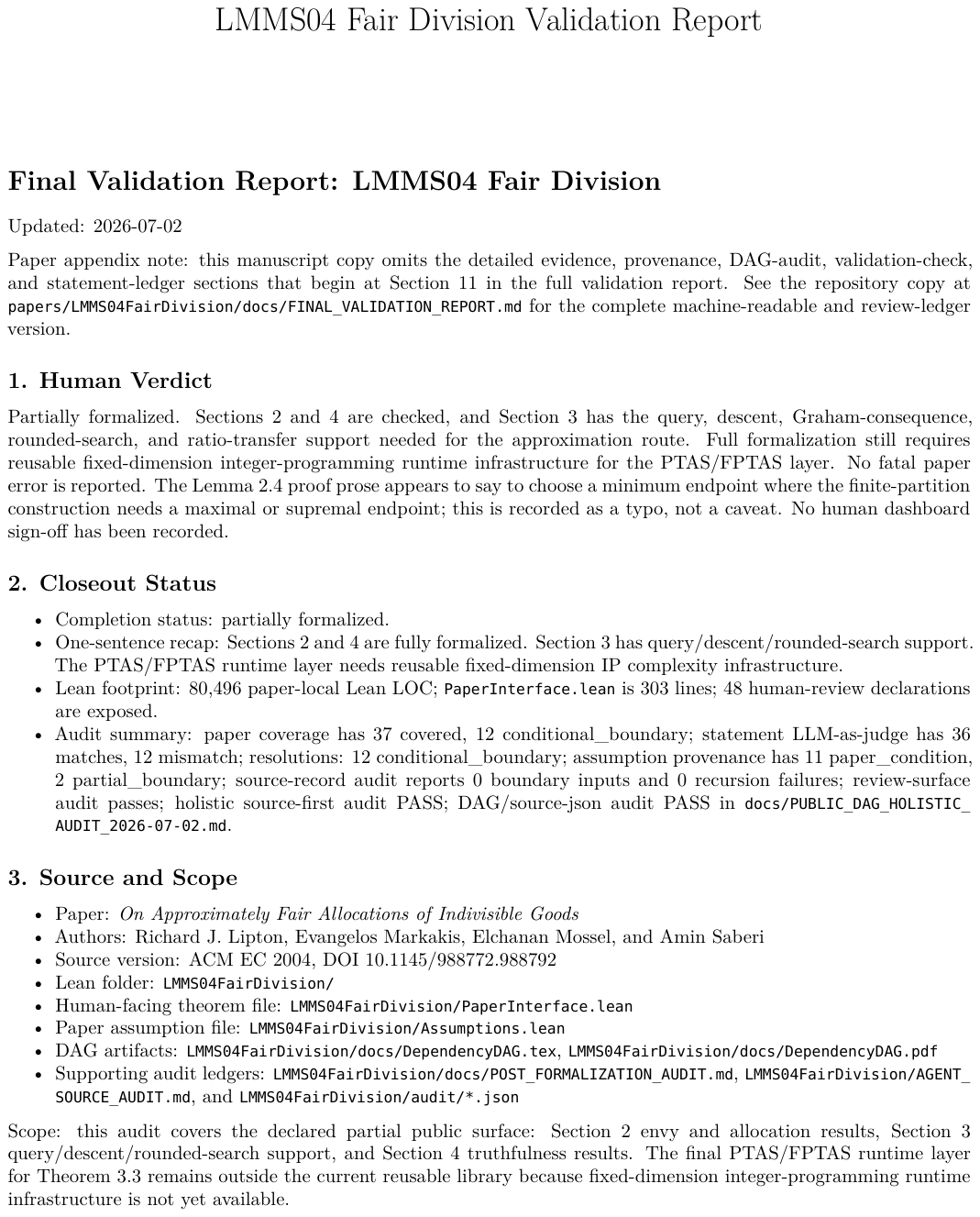}

\includepdf[pages=-,pagecommand={}]{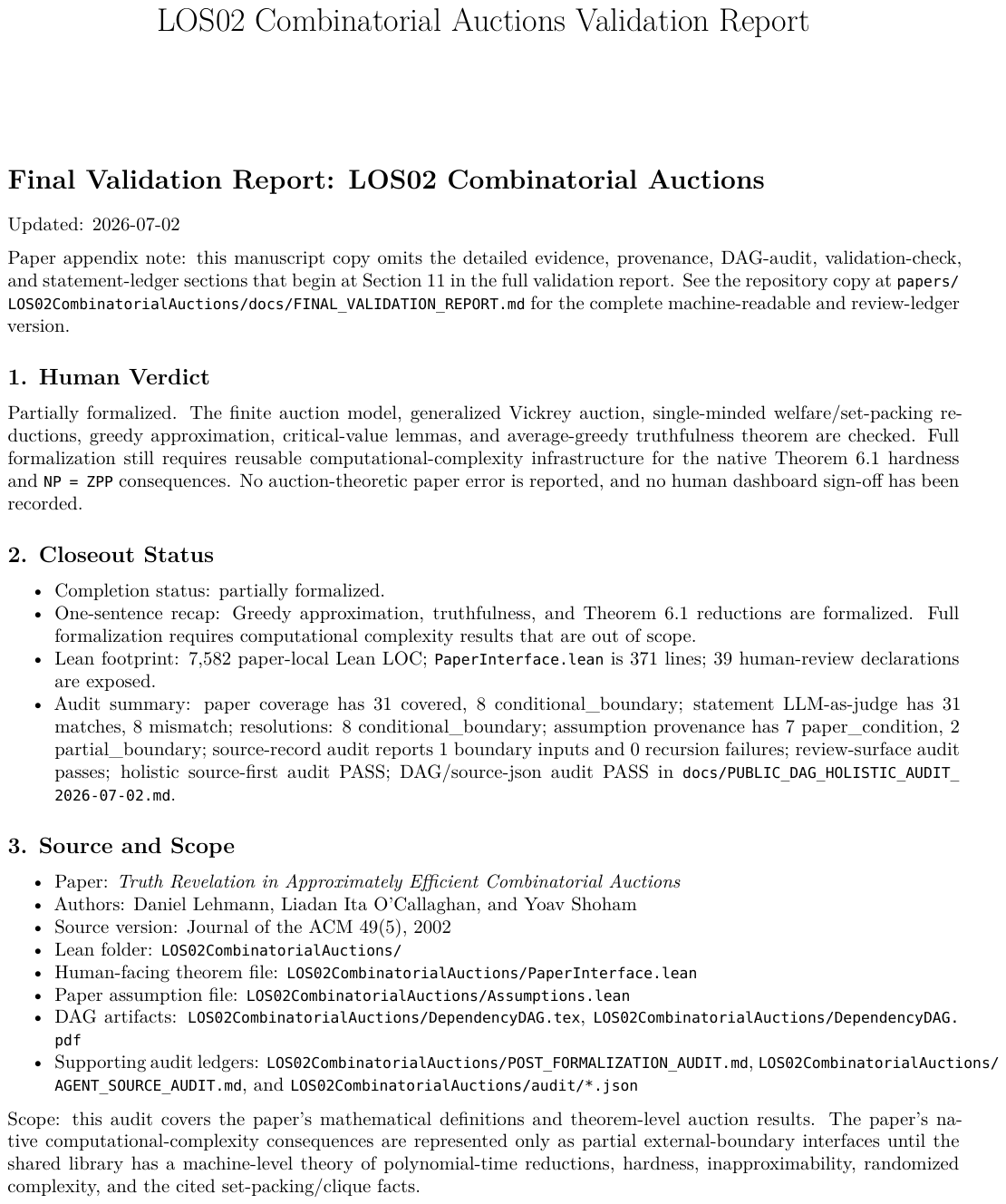}

\end{document}